\documentclass[prl,twocolumn,showpacs,superscriptaddress,lengthcheck]{revtex4}

\usepackage{amsmath,amssymb,amsthm}
\usepackage{graphicx}
\usepackage{color}
\usepackage{bbm}
\usepackage{subfloat}

\newcommand{\id}{\mathbbm{1}}

\begin{document}
\bibliographystyle{apsrev}

\title{Unifying variational methods for simulating quantum many-body systems}
\author{C.M.\ Dawson}

\affiliation{Blackett Laboratory, Imperial College London, Prince
Consort Road, London SW7 2BW, UK}

\affiliation{Institute for
Mathematical Sciences, Imperial College London, Prince's Gate,
London SW7 2PE, UK}

\author{J.\ Eisert}

\affiliation{Blackett Laboratory, Imperial College London, Prince
Consort Road, London SW7 2BW, UK}

\affiliation{Institute for
Mathematical Sciences, Imperial College London, Prince's Gate,
London SW7 2PE, UK}

\author{T.J.\ Osborne}

\affiliation{Department of Mathematics, Royal Holloway University of
London, Egham, Surrey TW20 0EX, UK}

\date{\today}

\begin{abstract}
We introduce a unified formulation of variational methods for simulating ground state properties of quantum many-body systems. The key feature is a novel variational method over quantum circuits via infinitesimal unitary transformations, inspired by flow equation methods. Variational classes are represented as efficiently contractible unitary networks, including the matrix-product states of density matrix renormalization, multiscale entanglement renormalization (MERA) states, weighted graph states, and quantum cellular automata. In particular, this provides a tool for varying over classes of states, such as MERA, for which so far no efficient way of variation has been known. The scheme is flexible when it comes to hybridizing methods or formulating new ones. We demonstrate the functioning by numerical implementations of MERA, matrix-product states, and a new variational set on benchmarks.
\end{abstract}
\maketitle

Quantum many-body systems pose some of the most difficult challenges
in modern physics, and many examples remain inaccessible to
analysis. Of the many methods that have been devised as attempts to
meet these challenges, one of the most successful has been the
\emph{density matrix renormalization group} (DMRG) \cite{Wilson}.
The DMRG was originally conceived as a numerical technique for
iteratively constructing the ground or low-energy states of a
Hamiltonian, so that the system's Hilbert space is truncated and the
difficulties associated with exponentially increasing dimension are
avoided. A more recent interpretation of the DMRG has cast it as a
variational method over \emph{matrix product states}
\cite{FCS,Vidal,Frank,Finite,Scholl}, and this shift in emphasis has
stimulated much work on extending its applicability.

Matrix product states are expected to
provide good approximations to the ground states of one-dimensional non-critical systems \cite{Onedim},
however in other cases it is expected that
alternative variational sets will be required. Motivated by this, new
classes have been introduced, such
as \emph{projected entangled pair states} \cite{Frank} and
\emph{weighted graph states} \cite{WGS} for higher-dimensional lattices, while \emph{multi-scale
entanglement renormalization} (MERA) \cite{MERA}
and \emph{contractor renormalization} \cite{CORE} may be more
appropriate for critical systems.
At first sight it may appear that these numerical approaches to
quantum many-body systems have little in common with each other. Moreover,
the specification of a variational class is only a first step -- we also
require an effective method of finding the best
description of the system's ground or low-energy states within that class.

\begin{figure}[h]
\begin{center}%
\resizebox{.93\columnwidth}{!}{\includegraphics{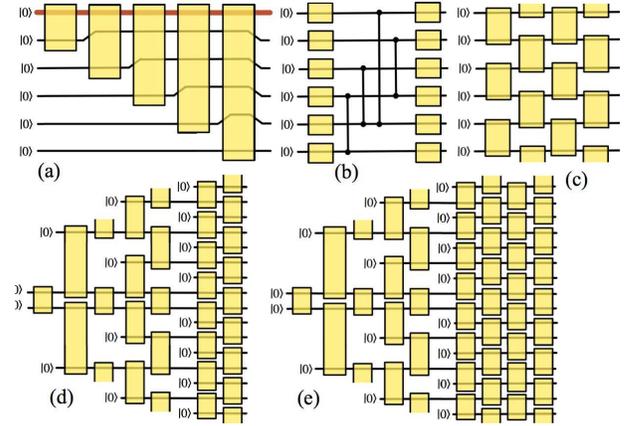}}%
\end{center}%
\caption{(a) An example of what is here called a
staircase circuit.
The upper system here is $d$-dimensional. Further examples of
unitary networks considered here include: (b) Unitary network of a
weighted graph state up to local deformations. (c)
Quantum cellular automata with Margolus partitioning. (d)
Quantum circuit of MERA including disentangling operations, (e)
A new variant combining (c) and (d) ({\it ``extended MERA''}).}
\label{fig:staircase}
\end{figure}

In this letter, we provide a unifying picture for several of these
variational methods: We show that by (i) recasting state classes as
quantum circuit classes (unitary networks) one can formulate a (ii)
general purpose variational method, related to the framework of {\it
flow equations} \cite{wegner94a}. 
We shall see that to provide a working
numerical method, it is sufficient that the propagation of each
Hamiltonian term can be efficiently tracked on a classical computer.
The \emph{contractibility} properties of the state classes mentioned
above translate immediately into an analogous property on their
corresponding circuits. Our approach -- a {\it flow equation
approach to variational simulations} -- may be regarded as an
optimal control approach \cite{wegner94a,Control} to varying
efficiently contractible networks describing variational states of
quantum many-body systems. It is flexible enough to hybridize known
methods or to construct new ones, and provides a first efficient way
of variation over MERA.

{\it Variational sets. --}
We begin with the casting of variational sets as unitary networks, which
provides the basis for the flow equations approach. Given $N$ spins
(and possibly ancillary systems) consider a family of states
    $S_d =\{
    U | \mathbf{0} \rangle : U\in {\cal U}_d
    \}$,
where $U\in {\cal U}_d$ is a set of unitary networks
characterized by some refinement parameter $d$,
and  $|\mathbf{0}\rangle$ denotes the state vector
with all spins down. The refinement
parameter plays the role, e.g., of the auxiliary dimension in
matrix-product states (MPS). These networks
consisting of gates $U=\prod_{j=1}^M U_j$
have to satisfy the condition that correlators of the form
  $ \langle \mathbf{0} |U^\dagger  O_1\dots  O_k U|\mathbf{0}
    \rangle$
can be efficiently computed for any $k$, that is with effort polynomial
in $N$ and  $d$.
Before turning to the variational method we will discuss a number of important
examples of states that can be discussed in this framework. We begin by
considering the matrix-product states of DMRG. For a system consisting of $N$
spins, we introduce a $d$-dimensional ancilla system, and consider a
circuit consisting of $N$ gates $U_j$ which act on both the ancilla system and qubit $j$,
giving a {\it ``staircase'' form}, see 
Fig.~\ref{fig:staircase}(a). By projecting the
output of such circuits onto
some basis state of the ancillary system or by disentangling it, we obtain any
matrix-product state on the spins. 
A second important example is the MERA class, see
Fig.~\ref{fig:staircase}(d).
This circuit is arranged in a tree 
structure with $\log N$ distinct \emph{layers}, each
of which introduces new spins into the circuit via two sets of gates known
as isometries and disentanglers
\cite{MERA}. An analogous circuit
is also possible using 2D binary trees. Further examples of quantum circuit classes
are weighted graph states,
where the refinement parameter $d$ is defined by the non-zero entries of the
adjacency matrix of the weighted graph, and {\it quantum cellular automata}
\cite{QCA}, the finite depth $d$ being the refinement parameter, and {\it new variants} as depicted in Fig.\ \ref{fig:staircase}.

{\it Flow equations as a unifying method of variation. --}
Before we introduce the method
of variation, let us first remind ourselves of flow equation ideas.
Consider a continuous transformation of an initial Hamiltonian $H$
    $H(t) = U(t)^\dagger H U(t)$,
where $U(t)$ is defined via a Hermitian generator
as the time-ordered integral $U(t) = \mathcal{T} \exp
(-i \int_0^t G(s) ds)$. The derivative of $H(t)$ is given by
  $ \label{eq:dynamical_eqn}
   \partial_t H = -i [G(t),H(t)]$.
A familiar example from optimal control theory chooses
$G(t) = i[K,H(t)]$, where $K$ is a real diagonal matrix with unique entries.
In this case $H(t)$ will converge to a diagonal matrix of
eigenvalues as $t\rightarrow\infty$, with the columns $U(t)$ the
corresponding eigenvectors. This is often referred to as
{\it double-bracket flow} \cite{wegner94a}. 
A straightforward application to quantum
many-body systems is impractical, as the flow will in general transform the
Hamiltonian into one having exponentially many terms.
The key to these methods is to 
{\it truncate} the resulting systems of differential
equations in a perturbative fashion that is a good approximation 
for small perturbations.

However, we are not aiming for approximate analytical expressions here. Consider a
quantum circuit as be a sequence of $M$ gates $U_j(t)$,
each of which is continuously parameterized with infinitesimal generator
$G_j(t)$ beginning with some arbitrary $U_j(0)$. Write
the overall unitary implemented by the circuit as
$U(t) = \prod_{j=1}^M U_j(t)$, and consider the
expectation of some many-body Hamiltonian
%\begin{equation}
%\label{eq:objective}
    $E(t) = \langle \mathbf{0}| U(t)^\dagger H U(t)
     |\mathbf{0}\rangle$.
%\end{equation}
A circuit class is defined here by a specification of the locations
of each gate $U_j(t)$, and the best approximation to a ground state
within a given class is the circuit that minimizes the expectation
$E(t)$. Within the framework of flow equations, we will show how one
can choose \emph{optimal} generators $G_j(t)$ for each gate.
Differentiating the expectation we get $\partial_t E  =
2\mathfrak{Re} \langle \mathbf{0} | U^\dagger H\partial_t U |
\mathbf{0} \rangle$, and our first goal is to minimize this
derivative subject to the Hilbert-Schmidt constraints
$\text{tr}[{G_j(t)^\dagger G_j(t)}] = \varepsilon$, i.e., the
generators remain ``infinitesimal''. For $U(t) = \prod_{j=1}^{M}
U_j(t)$, we find
\begin{eqnarray}
\label{eq:dU}
\partial_t U
& = & -i \sum_{j=1}^{M}\biggl(\prod_{k=j+1}^M U_k\biggr) G_j\biggl(\prod_{k=1}^{j} U_k\biggr).
\end{eqnarray}
Note the convention whereby $\prod_{j=1}^M U_j$ is ordered as $U_M \cdots U_2 U_1$, and the other way round for $\prod_{j=M}^1 U_j$.
We can substitute this back and minimize on a term-by-term basis at each
point $t$ of the flow.
Let $\{B^b\}$ be an appropriate orthonormal Hermitian operator basis,
and expand the $j$-th generator as
$G_j(t) = \sum_b g_{j,b} B^b$,
with $g_{j,b} $ real. Now define,
for the given Hamiltonian $H$,
\begin{equation}
\label{eq:defGamma}
    \Gamma_{j,b}(t) = \langle \mathbf{0} | U^\dagger H  \biggl(
    \prod_{k=j+1}^M U_k \biggr) B^b
    \biggl(\prod_{k=1}^{j} U_k \biggr)
| \mathbf{0} \rangle .
\end{equation}
Each term of the derivative with this parametrization is
$[\partial_t E ]_j = 2 \sum_b g_{j,b} \mathfrak{Re} [ \Gamma_{j,b}(t)]$,
and the constraints of the minimization problem are $\sum_b g_{j,b}^2 = \varepsilon$.
The Lagrange multiplier condition for a minimum is then simply
   $ g_{j,b} = -2\mathfrak{Re} [\Gamma_{j,b}(t) ]/\lambda$.
The following method of evaluating the optimal generator avoids
calculating the quantities $\Gamma_{j,b}(t)$ for each basis element
$B^b$: Writing $G_j$ out in its operator basis we have
   $ G_j(t)  = -({2}/{\lambda}) \sum_b \mathfrak{Re}
    \left[\Gamma_{j,b}(t)\right] B^b.$
Recall that the real part appears because we 
are taking an expectation of an operator and its
Hermitian conjugate. It will be convenient to set
$G_j = -({2}/{\lambda}) (F_j + F_j^\dagger )$, with
\begin{equation*}
\label{eq:fjtrace}
    F_j  =  \text{tr}\Big[\Big(\prod_{k=1}^{j} U_k \Big)
     |\mathbf{0}\rangle\langle\mathbf{0} | U^\dagger H
     \Big(\prod_{k=j+1}^M U_k\Big) B^b_j\Big] B^b_j,
\end{equation*}
%Within the trace we are implicitly considering each operator as acting on the entire Hilbert space,
%while the operator $F_j$ itself acts only a particular subsystem. 
%This may be written
which can after some steps be written
in terms of a {\it partial trace} over the subsystems $R_j$
not acted on by the gate $U_j$
\begin{equation}
\label{eq:gen_formula}
    F_j = \text{tr}_{R_j} \Big[
    \Big(\prod_{k=1}^{j} U_k \Big) |
    \mathbf{0}\rangle\langle\mathbf{0} |
    U^\dagger H \Big(\prod_{k=j+1}^M U_k\Big) \Big].
\end{equation}
The utility of this expression depends on whether or not we are able to efficiently calculate the partial trace, which
depends on the structure of the circuit class $\prod_{j=1}^M U_j$ and on the Hamiltonian $H$. Accordingly, we now introduce contraction techniques able to cope with such expressions \cite{FlowStrength}.

% -----------------------------------------------------
\begin{figure}[h]
\begin{center}
\resizebox{.92\columnwidth}{!}{\includegraphics{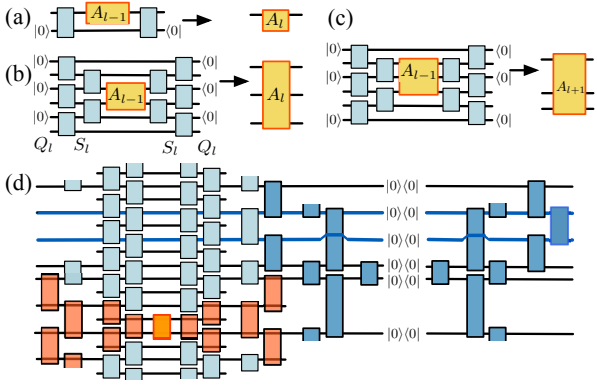}} \\ \vspace{1.2em}
\end{center}
\caption{Contraction rules that are used in the procedure of finding the
optimal generators (a-c), and a circuit correponding to  Eq.~(\ref{eq:gen_formula}) for
MERA (d). The red and blue shading indicate the two causal cones
that are encountered when evaluating the partial trace. \label{fig:typical_term}}
 \end{figure}
% -----------------------------------------------------

{\it Contraction. --}
We begin by reviewing some of the basic ideas of contraction and introduce a language for describing them. What we here call a standard contraction problem calls for the evaluation of an expectation of the form $\langle 0| U^\dagger A U |0\rangle$,
$U$ being a quantum circuit $\prod_{j=1}^M U_j$ acting on
$N$ spins (and any ancillae), and $A$ some observable.
A contraction procedure is a sequence of maps that construct operators $A_0, A_1, \ldots, A_L$ with $A_0 = A$ and
\begin{equation*}
A_{l} = \big(\langle {\mathbf 0}|_{Q_l}\big) \otimes \id  \big) \bigg(\prod_{j \in S_l} U_j\!\bigg)^\dagger \!\! A_{l-1} \bigg(\prod_{j \in S_l} U_j\!\bigg)   \big(|{\mathbf 0}\rangle_{Q_l} \otimes \id \big).
\end{equation*}
Here $Q_l$ and $S_l$ denote subsystems and subcircuits respectively, and are chosen so that at the final step we obtain $\langle {\mathbf 0}|A_L|{\mathbf 0}\rangle = \langle {\mathbf 0}| U^\dagger A U |{\mathbf 0}\rangle$.
The contraction is said to be efficient if the dimensions of the operators $\prod_{k \in S_j} U_k$ scale at most polynomially in the number of spins $N$.

The key point here is that we may evaluate such an expectation
(or similarly a trace or partial trace) without ever having to deal with the
overall unitary $U$, whose dimension is in general exponential in $N$.
As an example, suppose we have a two-body Hamiltonian term
$H_{j,j+1}$, and wish to evaluate $\langle 0|U^\dagger H_{k,k+1}
U|0\rangle$ for $U$ a staircase circuit. Then we set $A_0 = H_{k,k+1}$ and iterate
$A_{l} = \left(\id \otimes \langle 0|_{k-l}\right)U_{k-l}^\dagger A_{l-1} U_{k-l}\left(\id \otimes | 0\rangle_{k-l}\right)$,
so $Q_l$ here is simply the $l$th spin, and $S_l$ contains only the gate $U_l$. The final operator $A_L$ so obtained acts only on the ancilla system, and the desired expectation is thus $\langle 0 | A_L |0\rangle$.
At no stage in the procedure are we required to manipulate operators of dimension greater than
$2d \times 2d$. To be entirely clear, a representative of each of these
steps is depicted in Fig.\ \ref{fig:typical_term} (a).
A second example is provided by MERA circuits \cite{MERA}, which require
that we manipulate operators of
dimension at most $64 \times 64$.
Here the sets $Q_l$, $S_l$ are defined with respect to levels of the MERA
circuit and the
\emph{causal cone} of the given Hamiltonian term. A first such
step is represented in Fig.\ \ref{fig:typical_term} (b). After constructing
$A_l$, we can move on to the next MERA layer (Fig.\ \ref{fig:typical_term} (c)).

We now describe the
new contraction procedure used to evaluate
the more general
expressions of Eq.~(\ref{eq:gen_formula}), 
see Fig.\ \ref{fig:typical_term}. There are two sets of gates highlighted in these circuits, which will be dealt with in two separate contraction sequences, and we shall refer to these sets as the \emph{red cone} (originating from the Hamiltonian as explained above)
and \emph{blue cone} (originating from the generator) respectively. The
unhighlighted gates are simply cancelled as $U_k^\dagger U_k = \id$.
We can hence restrict ourselves to the causal cones. We shall also
refer to the gate $U_j$ whose generator we are calculating as the
\emph{generatee}.

The first contraction sequence involves the gates in the red cone. Here we set $A_0 = H_{j,j+1}$ and proceed via a sequence of partial expectations, with the same sets $Q_l$, $S_l$ used in a standard contraction (see again Fig.\ \ref{fig:typical_term} (b,c)).
This continues until we reach the set $S_G$ containing the generatee $U_j$.
At this point we are unable to continue as some or all of the gates in $S_G$
will have been cycled to the right-hand side (as we are calculating a partial trace).
The final operator of this contraction $A_G$ is then constructed by conjugating the previous $A_{G-1}$ by those gates that have \emph{not} been cycled.
The second contraction focuses on the blue cone, and begins by initializing $B_0 = |\mathbf{0}\rangle\langle\mathbf{0}|$, acting on the subsystem as the generatee. The contraction then iterates in the \emph{reverse} order to the standard contraction,
\begin{equation*}
B_{j} =\text{tr}\Big[{Q^\prime_j}{\Big(\prod_{k \in S^\prime_j} U_k\!\Big)  \big(B_{j-1} \otimes |\mathbf{0}\rangle\langle \mathbf{0}| \big) \Big(\prod_{k \in S^\prime_j} U_k\!\Big)^\dagger\Big]},
\end{equation*}
the primed $Q^\prime_j, S^\prime_j$ indicating the reversed order. This contraction also continues until it reaches the set $S_G$, at which point $B_G$ is constructed by (anti) conjugating $B_{G-1}$ with the gates from $S_G$ that \emph{have} been cycled. The operator $F_j$ is then given by $F_j = \text{tr}_{R_j} [   B_L A_L ]$, with $R_j$ as for Eq.~(\ref{eq:gen_formula}). For clarity, 
a MERA procedure is shown in
Fig.\ \ref{fig:typical_term} (d).

If the standard contraction procedure is efficient, then this
modified procedure will also be efficient as the largest operators
we must manipulate are defined by the same sets $S_l$. For example,
determining the optimal generator for a gate in a staircase circuit
acting on $N$ spins with ancilla dimension $d$ requires a time $O(N
d^3 )$, while for MERA the time required is $O(N \log N )$. The
above methods can be readily applied to 2D settings of MERA
\cite{MERA}, where one has, e.g., layers of Margolus
partitionings as in a quantum cellular automaton \cite{QCA}, with a
tree-like reduction of the number of sites in every second step;
then again, the contraction of the two cones can be done efficiently.

\begin{figure}[t]
\begin{center}
\resizebox{1.02\columnwidth}{!}{\includegraphics{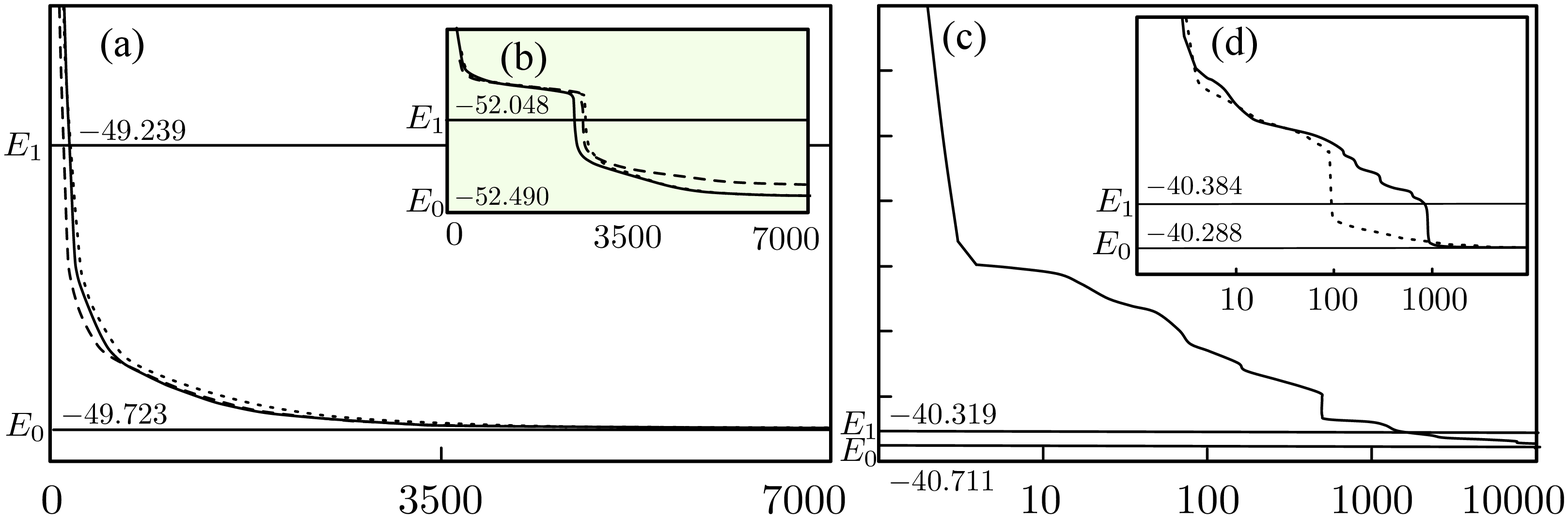}}
\caption{(a) Staircase flow for the $30$-qubit Heisenberg chain and
(b) $28$-qubit ring, in the number of flow steps. 
Plot (a) shows the decrease of expectation with the flow for $d = 10$ (dashed), $20$ (solid), $40$ (dotted). 
Exact values for the two lowest eigenvalues (for $N=30$) 
are also shown. (c) MERA flow for
the $32$-qubit Ising chain. (d) MERA flow for the
$32$-qubit Ising ring and extended MERA.}
\label{fig:implement}
\end{center}
\end{figure}

{\it Implementation. --}
We now have the main ingredients for an actual
algorithm. Fig.~\ref{fig:implement} illustrates
example implementations for the Heisenberg
and critical Ising Hamiltonians
\begin{eqnarray*}
    H_H= -\frac{J}{2}
    \sum_{j=1}^N  \sum_{k=1}^3 \sigma_j^k \sigma_{j+1}^k ,
    H_I= -\frac{1}{2}
    \sum_{j=1}^N  (\sigma_j^1 \sigma_{j+1}^1 + \sigma_j^3)
\end{eqnarray*}
as benchmarks. (a) and (b) illustrate an implementation for staircase
circuits for the Heisenberg chain (with both open and closed
boundary conditions) chosen for a first implementation as the
corresponding ALPS-DMRG provides good benchmark. Each step of the
flow requires the calculation and application of the optimal
generator for each gate. We see that for open boundary conditions
the staircase achieves, for the energy $E$, the same accuracy of
$\Delta=(E- E_0)/E_0$ as the benchmark ALPS-DMRG to
six significant digits, and no problems with local optima have been
observed \cite{Optima}. (c) presents a MERA implementation
(representative when random initial conditions are drawn) for the
$32$-qubit critical Ising model for bond dimension $2$. 
Even for this small bond dimension, the relative error of $\Delta=
4.4696\times 10^{-4}$ is achieved (note that this involves merely
$61$ unitaries acting on two spins, which is comparable in accuracy
to DMRG for a dimension $d$ defining MPS being described
by an order of magnitude of more real parameters), 
for the ring $\Delta=1.2901\times 10^{-4}$. Similar performance
is found for a $64$ spin model. (d) For the extended MERA we find comparable performance but quicker convergence. We have also
systematically compared the achievable accuracy for the Heisenberg
model with MERA with bond dimension $2$ (for which MERA performs
slightly worse) with the one extended MERA where one appends an
additional single layer of a quantum cellular automaton: This leads
in instances to a significant improvement (of the order of
$(E_1-E_0)/E_0$ in this model, critical in the
thermodynamical limit), but, first and foremost, 
shows the flexibility of the
approach \cite{Snake}.

{\it Conclusions and further work. --} We have shown how
ideas from flow equations may be adapted to provide variational
methods for approximating ground states of quantum
many-body systems. The appeal of this approach is its
flexibility as it is able to unify several existing ansatz classes
within a single framework with a universal variational
technique. Recent work into the dissemination of correlations
in quantum many-body systems has stimulated much work
on the construction of suitable variational classes. It is
hoped that the methods presented here will facilitate the
systematic exploration of the potential of these approaches.

{\it Acknowledgments. --}
This work
was supported by the DFG (SPP 1116), the EU (QAP), the EPSRC,
QIP-IRC, Microsoft Research, EURYI, and the Nuffield foundation.
We thank F.\ Verstraete and U.\ Schollw{\"o}ck for discussions.
Note added: For related work on MERA simulations
that has appeared on the preprint server in the meantime, see Refs.\ \cite{MT}.


\begin{thebibliography}{99}
\bibitem{FlowStrength}
    Eq.~(\ref{eq:gen_formula}) provides us with the optimal generator at each flow point $t$, but
    not the optimal
    flow ``strength'' $2/\lambda$.
    Poor choices of $\lambda$ may result in slow convergence.
    In the illustration presented below, 
    $E(t) = \langle \mathbf{0}| U(t)^\dagger H U(t)
     |\mathbf{0}\rangle$
    was well approximated, in the neighborhood
    of the optimum, by a quadratic in $1/\lambda$ and heuristic
    techniques can make use of this to determine a good
    flow strength. 
    The use of alternative techniques such
    as conjugate gradients or BFGS may be advantageous.
\bibitem{Snake}
    We have also used a 1D-MERA with bond dimension $2$
    in a 2D Ising model, in
    an ``snake-like'' ordering of the spins. The achieved
    accuracy -- although using essentially a 1D approach --
    is quite remarkable. In a $4\times4$ system
    the achieved accuracy is
    $\Delta= 1.5513\times 10^{-3}$,
    compared to DMRG energy of $\Delta=9.4412\times 10^{-2}$.
    For larger systems,
    1D-MERA outperforms 1D-DMRG even more distinctly.
\end{thebibliography}
\end{document}